\newcommand{\CC}{\hbox{{$\cal C$}}}
\newcommand{\R}{\mathbb{R}}
\newcommand{\C}{\mathbb{C}}
\newcommand{\Z}{\mathbb{Z}}

\newcommand{\Dsl}{{D\!\!\!\!/}}
\newcommand{\h}{{\scriptstyle\frac{1}{2}}}
\newcommand{\extd}{{\rm d}}
\newcommand{\del}{\partial}

\newcommand{\sgn}{{\rm sign}}

\newcommand{\eps}{{\epsilon}}
\newcommand{\tens}{\mathop{\otimes}}

\newcommand{\Ad}{{\rm Ad}}
\newcommand{\id}{{\rm id}}

\renewcommand{\>}{\rangle}

\newcommand{\proof}{{\bf Proof\ }}
\newcommand{\eproof}{$\quad \diamond$\bigskip}

\newcommand{\eqn}[2]{\begin{equation}#2\label{#1}\end{equation}}

\documentclass[11pt]{article}
\usepackage{amssymb,amsmath,epsfig}
\newtheorem{lemma}{Lemma}[section]
\newtheorem{propos}[lemma]{Proposition}

\newtheorem{theorem}[lemma]{Theorem}

\begin{document}\baselineskip 18pt

{\ }\qquad \hskip 4.3in
\vspace{.2in}

\begin{center} {\LARGE ELECTROMAGNETISM AND GAUGE THEORY ON THE
PERMUTATION GROUP $S_3$} \\ \baselineskip 13pt{\ }\\
{\ }\\ Shahn Majid and E. Raineri \\ {\ }\\ School of Mathematical
Sciences, Queen Mary and Westfield College\\ University of
London, Mile End Rd, London E1 4NS, UK
\end{center}
\begin{center}
December, 2000 -- revised
\end{center}
%\vspace{10pt}

\begin{quote}\baselineskip 13pt
\noindent{\bf Abstract} Using noncommutative geometry we do U(1)
gauge theory on the permutation group $S_3$. Unlike usual lattice
gauge theories the use of a nonAbelian group here as spacetime
corresponds to a background Riemannian curvature. In this
background we solve spin 0, 1/2 and spin 1 equations of motion,
including the spin 1 or `photon' case in the presence of sources,
i.e. a theory of classical electromagnetism. Moreover, we solve
the U(1) Yang-Mills theory (this differs from the U(1) Maxwell
theory in noncommutative geometry), including the moduli spaces
of flat connections. We show that the Yang-Mills action has a
simple form in terms of Wilson loops in the permutation group,
and we discuss aspects of the quantum theory.

 \end{quote}

\baselineskip 18pt
\section{Introduction}

As an attempt to make quantum theory computable it is common to
consider its formulation on a flat lattice $\Z^n$ in place of
spacetime $\R^n$. On the other hand, using modern methods of
noncommutative geometry it is possible to formulate such
constructions more `geometrically' in terms of a noncommutative
exterior algebra of differential forms and a Cartan calculus. In
lattice approximations the finite-differences are indeed
intrinsically noncommutative in the sense that they should be
formulated better as bimodules over functions: the product of a
function and a finite differential is naturally given by the
value of the function either at the start-point or the end-point
of the differential, and the two are different. Hence functions
and 1-forms obey $f \extd x \ne (\extd x)f$, which means that
such a more general noncommutative geometry is the natural way to
do lattice theory.

In this paper we want to go much beyond this initial observation.
In fact such methods of noncommutative geometry apply equally
well for any Hopf algebra and hence in particular for any finite
group $G$. This offers the possibility for the first time of a
natural `geometric' lattice approximation by nonAbelian and
finite groups rather than by a $\Z^n$ lattice. Using periodic
boundary conditions it is of course not hard to do lattice
computations on finite cyclic groups, although even this case is
interesting in noncommutative geometry\cite{MaSch:lat}. However,
the noncommutative theory comes into its own when we seek to model
for example a space or spacetime with spherical or other
topology. In particular it has been shown recently in
\cite{Ma:fin} that just as cyclic groups $\Z_n$ approximate tori,
permutation groups such as $S_3$ (permutations on 3 elements) are
more like compact semisimple Lie groups. It was shown for example
that $S_3$ has a natural noncommutative Riemannian structure with
Ricci curvature essentially proportional to the metric and
translation-invariant (like a classical sphere $S^3$).  The
curvature originates in the nonAbelianess of the group $S_3$.

Other metrics and connections also exist and on principle one
could proceed to gravity and quantum gravity on $S_3$ using these
methods. Before attempting such a project one should consider the
simpler problem of spin 0, 1/2, 1 fields moving in the natural
Killing-form metric Riemannian background.This is what we do in
the present paper. In the natural 3-bein coordinates the Killing
metric just turns out to be\cite{Ma:fin} the Euclidean
$\delta_{ab}$. Using this we then define the Hodge $\star$
operator and hence such things as the Maxwell and Yang-Mills
Lagrangians $(\extd F)^*\wedge\star F$. The classical theory
particularly of `electromagnetism' explores in effect the
classical noncommutative geometry of $S_3$. We compute the
quantum deRham cohomology (it is nontrivial) and linear wave
equations etc. in Section~2. We also obtain point sources and
dipole sources for the Maxwell field.
We also explain the required Coulomb gauge fixing and more
or less completely treat the linear system.

In Section~3 we look at the non-linear $U(1)$ Yang-Mills theory
with $F=\extd A+A\wedge A$ (this is not the same as the linearised Maxwell
theory due to the non(super)commutativity of the differential forms).
We find the moduli space of flat connections, which turns out to be
nontrivial. We also look for instantons but show that none exist obeying the
required reality conditions. Finally, we show that the Lagrangian in the
Yang-Mills case has a nice description in terms of a real `kinetic' term and
Wilson loops around elementary plaquettes,
\[ L=\lambda^2_u\del^u\lambda^2_v+\lambda^2_u\lambda^2_v
-W_u(A)+{\rm cyclic\ rotations}\] where $u,v,w$ are the
transpositions of $S_3$ and label the tangent space at each point
$x\in S_3$, $\lambda_u$ etc are real positive fields built from
$A$ (essentially we use polar coordinates for the values of $A$)
and
\[ W_u(A)(x)=(1+A^u(x))(1+A^v(xu))(1+A^u(xuv))(1+A^w(xw))\]
is the holonomy around a small square at $x$ with sides $u,v,u,w$
in the group. It is remarkable that we do not put this in by hand
as some kind of approximation (as one does in conventional lattice
theory), it is literally what we
obtain for $F^*\wedge\star F$ using the noncommutative
differential geometry on $S_3$ and the Riemannian structure from
\cite{Ma:fin}. The latter, however, works equally for essentially
all quantum groups and many other systems, though we do not
discuss them here.

We conclude in Section~4 with some remarks about the quantum
theory. There being only six points in $S_3$ functional integrals
over our fields become multiple usual integrals. We formulate the
required actions based on minimal coupling and also explain how to
compute the partition function and expectation values of Wilson
loops $\<W_u(A)(x)\>$. All of this should be viewed as a warm up
to functional integrals over metrics and their connections i.e.
quantum gravity where our finite method should be particularly
useful. An introduction to the framework of gravity in our
approach (which plays only a background role) is in
\cite{Ma:cons}.

\subsection{Preliminaries}

Here we recall very briefly the formalism of noncommutative
differential geometry for finite groups $G$. This Hopf algebra
approach to noncommutative geometry coming out of quantum groups
should not be confused with the approaches to noncommutative
geometry of Connes\cite{Con:geo}, though the treatment of
one-forms as bimodules is common to both, and there are some
models where the two methods begin nontrivially to
`converge'\cite{MaSch:lat}.

In the quantum groups approach we work with the algebra $\C[G]$ of
functions on $G$. We do {\em not} consider derivations as vector
fields (this does not work here) but rather we define $\Omega(G)$
the exterior algebra of forms as a $\Z_2$-graded algebra with
$\extd$ a super-derivation and $\extd^2=0$. Using the
construction of \cite{Wor:dif} this is specified in a bicovariant
manner entirely by an $\Ad$-stable subset $\CC$ not containing
the group identity $e$. The one forms have a basis $\{e_a:\
a\in\CC\}$ over $\C[G]$, bimodule structure and $\extd$ on
functions
\[ \Omega^1=<e_a>, \quad e_a f=R_a(f)e_a,\quad \extd f=\sum_a (\del^a
f) e_a,\quad \del^a=R_a-\id\] where $ R_a(f)(x)=f(xa)$ for all
$x\in G$ and $a\in \CC$. The elements of $\CC$ are the `allowed
directions'. The partial derivatives defined here obey a
braided\cite{Ma:cla} Leibniz rule
\[ \del^a(fg)=\del^a(f)g+R_a(f)\del^a(g),\quad \forall f,g\in\C[G].\]
The higher forms are a certain quotient of the
tensor power of 1-forms where we set to zero those `symmetric'
combinations invariant under a braided-symmetrization operator
defined by a certain braiding $\Psi$. The $\extd$ is extended
through the Maurer-Cartan relation
\[ \extd e_a=e_a\wedge\theta+\theta\wedge e_a,\quad \theta\equiv\sum_a
e_a\]and the graded Leibniz rule. From this one also finds that
\[ \extd\alpha=[\theta,\alpha\},\quad \forall \alpha\in \Omega(G)\]
using the graded anti-commutator. Also
\[ e_{a_1}\wedge\cdots \wedge e_{a_m} f=R_{a_1\cdots
a_m}(f)e_{a_1}\wedge \cdots\wedge e_{a_m}\] where the product
$a_1\cdots a_m$ defines a natural $G$-valued degree on
$\Omega(G)$. Further details of the set-up including the required
quotient at degree 2 for general $G$ are in \cite{Ma:fin}.

For $S_3$ we take generators and relations, and conjugacy class
\[ u^2=v^2=e,\quad uvu=vuv\equiv w,\quad \CC=\{u,v,w\}.\]
So $\Omega^1=<e_u,e_v,e_w>$. Because every element of $\CC$ has
order 2, we have
\[ R_a\del^a=-\del^a,\quad (\del^a)^2=-2\del^a\]
for all $a=u,v,w$. It is also easy to see that degree 2-relations
\[ e_u\wedge e_v+e_v\wedge e_w+e_w\wedge e_u=0,\quad
e_v\wedge e_u+e_w\wedge e_v+e_u\wedge e_w=0\]
\[ e_u\wedge e_u=e_v\wedge e_v=e_w\wedge e_w=0\]
hold. It is well-known that these are in fact the only relations
in degree two in the Woronowicz construction (an actual proof is
in \cite{Ma:fin}). Hence $\Omega^1$ is 3-dimensional while
$\Omega^2$ is 4-dimensional. As a basis of the latter we choose
(for concreteness)
\[ \Omega^2=<e_u\wedge e_v,\ e_v\wedge e_u,\ e_v\wedge e_w,\ e_w\wedge
e_v>.\]

Next, one easily computes the consequences of the degree 2
relations in higher degree, which we call the `quadratic
prolongation' of $\Omega^1$. It has been used for $S_3$
in \cite{Bre} and recently, for example, in \cite{Cas:gra} and one has
\[ e_u\wedge e_v\wedge e_w=e_w\wedge e_v\wedge e_u=-e_w\wedge
e_u\wedge e_w=-e_u\wedge e_w\wedge e_u\] and the two cyclic
rotations $u\to v\to w\to u$ of these relations. Hence there are
three independent 3-forms
\[ \Omega^3=<e_w\wedge e_u\wedge
e_v,\ e_u\wedge e_v\wedge e_w, e_v\wedge e_w\wedge e_u>\] in the
quadratic prolongation. Similarly there is one independent 4-form
with
\[ {\rm Top}\equiv e_u\wedge e_v\wedge e_u\wedge e_w=e_v\wedge e_u\wedge
e_v\wedge e_w=-e_w\wedge e_u\wedge e_v\wedge e_u=-e_w\wedge
e_v\wedge e_u\wedge e_v\] and equal to the 2 cyclic rotations of
these equations ({\rm Top} is invariant). Any expression of the
form $e_a\wedge e_b\wedge e_a\wedge e_b$ is zero as is any
expression with a repetition in the outer (or inner) two
positions. It is easy to see that the basic 2-forms mutually
commute and that $\rm Top$ has trivial total $G$-degree.

It turns out that {\em the quadratic prolongation in this case is
exactly $\Omega(S_3)$}, i.e. there are no further relations
imposed by the braided-antisymmetrization process in higher
degree in this case. This is {\em not} expected to hold in general
and we have not seen an actual proof of this fact for $S_3$,
therefore we include it now for completeness.

\begin{lemma} There are no further relations from Woronowicz's braided
antisymmetrization procedure, i.e. $\Omega(S_3)$ has dimensions
$1:3:4:3:1$ as for the quadratic prolongation.
\end{lemma}
\proof According to \cite{Wor:dif} we have to compute the dimension of
the kernel of
\[
A_3=\id-\Psi_{12}-\Psi_{23}+\Psi_{12}\Psi_{23}+\Psi_{23}\Psi_{12}
-\Psi_{12}\Psi_{23}\Psi_{12}\] acting on
$\Omega^1\tens\Omega^1\tens \Omega^1$ (tensor over $\C[S_3]$).
Here the braiding is $\Psi(e_a\tens e_b)=e_{aba^{-1}}\tens e_a$.
To find the dimension of the kernel, one first checks that
\(A_3(e_{a} \tens e_{b} \tens e_{c})=0 \) as soon as \(a=b\) or \(b=c\).
The null space of $A_3$ spanned by these vectors has a complement
\(V=\bigoplus_{c \in \CC}V_c \) where, for \(c \in \CC \)
and \((a,b,c)\) a cyclic permutation of \((u,v,w)\), \(V_c\) has basis
\[  \{ e_{a} \tens e_{b} \tens e_{a},\; e_{b} \tens e_{a} \tens e_{b},\;
e_{b} \tens e_{c} \tens e_{a},\; e_{a} \tens e_{c} \tens e_{b} \}
 \]
One finds that each \(V_{a}\) is preserved by \(A_3\), and \(A_3\) is
given by this \(4 \times 4 \) matrix (in the chosen basis)
\[
\left(\begin{matrix}
1 & 1 & -1 & -1 \\
1 & 1 & -1 & -1 \\
-1 & -1 & 1 & 1 \\
-1 & -1 & 1 & 1
\end{matrix}\right), \]
which is diagonalisable with  eigenvalues \((0,0,0,4)\).
Therefore \(\mbox{dim} A_3(V_c)=1\)
for all \( c \in \CC \), and
\(\dim (A_3 (\Omega^{1})^{\tens 3})=\sum_{a \in \CC} 1=3\).
Hence $\Omega^3$ which is defined as the tensor cube of $\Omega^1$
modulo
$\ker A_3$ is 3 dimensional, which is the same as the quadratic
prolongation, so that there are no further relations in degree 3.
Notice that \(A_3\) is not a projector, but
\(\frac{1}{4}A_3\) is.
\\
In degree 4 we check that $\rm Top$ is not in the kernel of $A_4$
(defined similarly) and hence that there is no further quotient
in degree 4.
\eproof

Next it is obvious in the presence of a {\rm Top} form that one
can define $e_a\wedge e_b\wedge e_c\wedge e_d=\eps_{abcd}{\rm
Top}$ for all $a,b,c\in\CC$. This is not yet enough to proceed in
to a Hodge $\star$ operator because for that one needs a
Riemannian metric $\eta_{ab}$. However, this is precisely what
comes out of the theory of Riemannian structures on finite groups
and quantum groups\cite{Ma:fin} from the `braided Killing form' of
the tangent space braided-Lie algebra. For $S_3$ (in a suitable
normalisation) it just turns out to be $\eta_{ab}=\delta_{ab}$,
the Euclidean metric in the natural 3-bein coordinates provided by
the $e_a$ themselves. Using this we now introduce the Hodge $*$
operator
\[ \star(e_{a_1}\wedge\cdots\wedge e_{a_m})=d_m^{-1} \eps_{a_1\cdots
a_mb_{m+1}\cdots b_n}\eta^{b_{m+1}c_{m+1}}\cdots
\eta^{b_nc_n}e_{c_n}\wedge \cdots\wedge e_{c_{m+1}}=d_m^{-1}
\eps_{a_1\cdots a_n}e_{a_n}\wedge \cdots\wedge e_{a_{m+1}}\] for
some normalisation constants $d_m$. The ordering of indices is
determined so that the total $G$-degree (as above) is preserved by
$\star$ (here every element of $\CC$ has order 2 or we would need
inverses on the right hand side).  In our case we take
\[ d_0=1,\quad d_1=2,\quad d_2=\sqrt{3},\quad d_3=2,\quad
d_4=1.\]
In this way one finds:

\begin{propos} The natural Hodge $\star$ operator on $\Omega(S_3)$ is
\[\star(1)={\rm Top},\quad  \star(e_u)=2 e_w\wedge e_u\wedge e_v,
\quad \star(e_v)=2 e_u\wedge
e_v\wedge e_w,\quad \star(e_w)=2 e_v\wedge e_w\wedge e_u\]
\[\star(e_u\wedge e_v)=-3^{-\h}(e_u\wedge e_v+2e_v\wedge e_w),\quad
\star(e_v\wedge e_w)=3^{-\h}(e_v\wedge e_w+2 e_u\wedge e_v)\]
\[ \star(e_v\wedge e_u)=3^{-\h}(e_v\wedge e_u+2e_w\wedge e_v),\quad
\star(e_w\wedge e_v)=-3^{-\h}(e_w\wedge e_v+2e_v\wedge e_u)\]
\[ \star(e_w\wedge e_u\wedge e_v)=-\h e_u,\quad
\star(e_u\wedge e_v\wedge e_w)=-\h e_v,\quad \star(e_v\wedge e_w\wedge
e_u)=-\h e_w,\quad \star{\rm Top}=-1\]
extended as a bimodule map. It obeys $\star^2=-\id$.
\end{propos}
\proof By its construction it is clear that $\star$ has square -1 and
preserves the $G$-degree. The latter means that if we define
$\star(fe_{a_1}\wedge \cdots \wedge
e_{a_m})=f\star(e_{a_1}\wedge\cdots\wedge e_{a_m})$ for any function
$f$ then also $\star(e_{a_1}\wedge\cdots\wedge
e_{a_m}f)=\star(R_{a_1\cdots a_m}(f)e_{a_1}\wedge\cdots\wedge e_{a_m})=
R_{a_1\cdots a_m}(f)\star(e_{a_1}\wedge\cdots\wedge
e_{a_m})=\star(e_{a_1}\wedge\cdots\wedge e_{a_m})f$ as required. Note
also that since $\rm Top$ is cyclically invariant there is also a cyclic
invariance of $\star$. \eproof

Also associated to this metric is a Riemannian covariant
derivative, spin connection and Dirac operator. We will need the
latter (coupled to a further U(1) gauge field) in later sections.
However, for spin 0,1 one may proceed with only the Hodge $\star$
as above. As far as we know this Riemannian and Hodge structure
goes beyond what has been considered before. Finally, whereas the
above results hold over any field of characteristic zero, we also
impose a complex $*$-algebra structure when we work over $\C$.
Thus we define
\[ e_a^*=e_a,\quad \extd (\alpha^*)=(-1)^{|\alpha|+1}(\extd \alpha)^*\]
and one may check that $\Omega(G)$ becomes a differential graded
$*$-algebra. This should not be confused with the Hodge operator above.

\section{Wave Equations on $S_3$}

In this section we write down Lagrangians and solve the
associated linear wave equations for different spin. The spin 1
case means here `Maxwell theory' or 1-forms modulo exact. This is
a linearized version of the noncommutative U(1) gauge theory in
Section~3.

\subsection{Spin 0}

We consider a scalar field $\phi\in \C[S_3]$. From the definitions,
\[(\extd\phi)^*=e_a^*\overline{\del^a\phi}=e_a\del^a\bar\phi
=R_a(\del^a\bar\phi)e_a=-\del^a\bar\phi
  e_a=-\extd\bar\phi\]
as it should, and also note that
\[ e_a\wedge\star(e_b)=2\delta_{a,b}{\rm Top}.\]
Hence
\[L{\rm Top}\equiv -\h  (\extd\phi)^*\wedge \star(\extd
\phi)=\h \sum_{a,b}(\del^a\bar\phi) e_a (\del^b\phi)
\star(e_b)=\sum_a (\del^a\bar\phi)R_a(\del^a\phi){\rm Top}
\] gives the Lagranian density as
\[ L=-\sum_a\del^a\bar\phi \del^a\phi\]
for scalar fields. Using the braided-Leibniz rule
this is up to a total derivative
\[ L=\sum_a (R_a\bar\phi)(\del^a)^2\phi=-\sum_aR_a(\bar\phi (\del^a)^2\phi)
=-\sum_a\bar\phi(\del^a)^2\phi\] again up to total derivatives.
Hence the wave operator on spin zero is
\[ \square =-\sum_a \del^a \del^a =\sum_a 2\del^a.\]
It is easy to solve this. On a group manifold we would expect
`plane waves' associated to irreducible representations.

\begin{propos} The only zero mode of $\square$ is the constant function.
In addition there is one mode of mass $2\sqrt{3}$ given by the
sign representation, and four modes of mass $\sqrt{6}$ given by the
matrix elements of the 2-dimensional representation of $S_3$.
\end{propos}
\proof  In our case $S_3$ has a trivial representation, which
gives $\phi=1$ with `mass' zero. Then it has the sign
representation which gives
\[ \phi(x)=\sgn(x)\equiv (-1)^{l(x)},\quad \square\phi(x)
=2\sum_a ((-1)^{l(xa)}-(-1)^{l(x)})
=-12\phi(x)\]
with `mass' $2\sqrt{3}$ (here $l(x)$ is the length of the
permutation or the number of $u,v$ in its reduced expression).
Finally it has a $2\times 2$ matrix representation
\[ \rho(u)=\left(\begin{matrix}0&1\\ 1&0\end{matrix}\right),\quad
\rho(v)=\left(\begin{matrix}1&0\\-1&-1\end{matrix}\right)\]
and each matrix element (for each $i,j=1,2$ fixed)
\[ \phi_{ij}(x)=\rho(x)^i{}_j\]
is a `mass' $\sqrt{6}$ since
\[\square\phi_{ij}(x)
=-6\phi_{ij}(x)+2\sum_a \sum_k
\rho(x)^i{}_k\rho(a)^k{}_j=-6\phi_{ij}(x)\] as
$\rho(u)+\rho(v)+\rho(w)=0$. These four waves are linearly
independent because the representation is irreducible. Since
$\square$ is a $6\times 6$ matrix we have completely diagonalised
it, i.e. its eigenvalues correspond to allowed masses
$0,2\sqrt{3},\sqrt{6}$ with multiplicities $1,1,4$. \eproof

Moreover, every function on $S_3$ has a unique decomposition of
the form
\[ \phi=p_0+p_1\sgn + p_{ij}\phi_{ij}\]
for some numbers $p_0,p_1,p_{ij}$ (real if we demand
$\bar\phi=\phi$) i.e. a sum of our six waves. Associated to this
decompositon is a projection of any function to its component
waves (or nonAbelian Fourier transform). It is also worth noting
that $\square$ is hermitian with respect to the usual $L^2$ inner
product on $S_3$ and bicovariant hence its eigenspace
decomposition must exist and be a decomposition into $S_3\times
S_3$ modules (similarly for any group $G$). In the $S_3$ case at
least it is precisely the Peter-Weyl decomposition obtained in a
new way.

We note that there is another useful construction of the
projection to the mass $\sqrt{6}$ part, namely let $\phi_0$ be any
function and consider
\[ \phi=2\phi_0-R_{uv}\phi_0-R_{vu}\phi_0\]
Then
\[ \square\phi=-6\phi+2\sum_a R_a(\phi)=-6\phi+2\sum_a
(2R_a\phi_0-R_{auv}\phi_0-R_{avu}\phi_0)=-6\phi\] so $\phi$ is a
solution of mass $\sqrt{6}$. One should divide by 3 for an actual
projection of course. One may similarly project onto the other
waves.

\subsection{Spin 1/2}

For uncharged spin 1/2 we use the `curved space' Dirac operator
introduced in \cite{Ma:fin}. There, the `gamma-matrices' are given
explicitly by

\[
\gamma_u={1\over 3}\left(\begin{matrix}-1&1\\
1&-1\end{matrix}\right), \quad \gamma_v={1\over
3}\left(\begin{matrix}0&0\\ -1&-2\end{matrix}\right), \quad
\gamma_w={1\over 3}\left(\begin{matrix}-2&-1\\
0&0\end{matrix}\right)\] and obey
\eqn{gamrelS3}{\gamma_a\gamma_b+\gamma_b\gamma_a+{2\over
3}(\gamma_a+\gamma_b)={1\over 3}(\delta_{ab}-1),\quad
\sum_a\gamma_a=-1.} There is a natural spin connection
corresponding to the Killing form metric on $S_3$ and including
this, one has\cite{Ma:fin},
\[ \Dsl=\del^a\gamma_a-1={1\over
3}\left(\begin{matrix}-\del^u-2\del^w
-3&\del^u-\del^w\\
\del^u-\del^v& -\del^u-2\del^v-3\end{matrix}\right)={1\over
3}\left(\begin{matrix}-R_u-2R_w& R_u-R_w\\
R_u-R_v& -R_u-2R_v\end{matrix}\right).\]

It acts on 2-vector valued functions (spinors) on $S_3$. We note
that if we let $\gamma=\sgn$ acting by pointwise multiplication
then
\[ \{\Dsl,\gamma\}=0.\]
This should not be viewed as chirality since it acts on the spinor
components as functions not on the spinor values. It does,
however, mean that solutions are paired with massive eigenvalue
$m$ going to eigenvalue $-m$ under $\gamma$. We define mass here
as the negative eigenvalue of $\Dsl$. Note, however, that $\Dsl^2$
is a second-order operator (it involves $R_{uv},R_{vu}$) and not
merely $\square$ plus a scalar curvature term as in the
Lichnerowicz formula.

\begin{propos}
$\Dsl$ has $4$ zero modes, four massive modes with eigenvalue $+1$
modes and four with eigenvalue $-1$ modes related by $\gamma$.
\end{propos}
\proof To find the solutions we consider first of all spinors of
the form
\[ \psi=\left(\begin{matrix}R_{uv}\phi\\
\phi\end{matrix}\right)\] for some function $\phi\in\C[S_3]$. The
Dirac operator reduces to
\[ \Dsl\psi=-{1\over 3}\sum_a R_a\psi=(-1-{1\over 6}\square)\psi\]
acting on each component. Hence there are 4 linearly independent
zero modes of the form
\[ \psi_{ij}=\left(\begin{matrix}R_{uv}\phi_{ij}\\
\phi_{ij}\end{matrix}\right)\] induced by the spin 0 waves
$\phi_{ij}$ of mass $\sqrt{6}$. We also have a massive mode of
eigenvalue $-1$ from $\phi=1$ and $+1$ from $\phi=\sgn$ from the
remaining spin 0 waves, but these solutions are obvious by
inspection. In fact it is obvious that
\[ \psi_+=\left(\begin{matrix}1\\0\end{matrix}\right),\quad
\psi_-=\left(\begin{matrix}0\\1\end{matrix}\right)\] are
separately solutions of eigenvalue $-1$, and similarly with $\sgn$
for eigenvalue $+1$.

Two further and independent solutions of eigenvalue $-1$ are
obtained by the similar ansatz
\[ \psi=\left(\begin{matrix}\phi\\
R_{uv}\phi\end{matrix}\right).\] This time
\[ \Dsl\psi=\left(\begin{matrix}\Delta\phi\\
R_{uv}\Delta\phi\end{matrix}\right),\quad \Delta=R_v-{2\over
3}\sum_a R_a\] which is easily solved by $\phi$ a linear
combination of the $\phi_{ij}$. The 2nd term of $\Delta$ vanishes
on these and $R_v\phi_{ij}=\phi_{ik}\rho(v)^k{}_j$. But $\rho(v)$
has precisely one eigenvector $\alpha$ of eigenvalue $-1$ and
hence contracting with this gives a pair of solutions
$\phi=\phi_{ij}\alpha^j$ of eigenvalue $-1$. In the basis used
above, the resulting two massive spinor waves of eigenvalue $-1$
are
\[ \psi_i=\left(\begin{matrix}\phi_{i2}\\
R_{uv}\phi_{i2}\end{matrix}\right).\] They are linear independent
since $\rho$ was irreducible. Similarly for eigenvalue $+1$ if we
use the $+1$ eigenvector of $\rho(v)$. Altogether we have a
complete diagonalisation of $\Dsl$. \eproof

 We can consider
real or complex spinors (in fact the linear theory works over any
field of characteristic zero). For a general group $G$ any
irreducible representation $\rho$ similarly defines
$\gamma_a$-matrices\cite{Ma:fin} and one can expect a similar
method to the above to diagonalise $\Dsl$, with mass spectrum
related to the eigenvalues of $\CC$ in the representation.

\subsection{Zero curvature Maxwell fields and deRham cohomology}

For a spin 1 or Maxwell `photon' field we take a 1-form
$A\in\Omega^1$ defined modulo exact differentials or `linearised
gauge transformations'. The well-defined curvature is of course
\eqn{Fda}{F=\extd A.} For example, the moduli space of flat
connections modulo gauge transformations in this linearised
context is the cohomology $H^1$ with respect to the noncommutative
differential forms.

\begin{propos} The noncommutative deRham cohomology of $S_3$ is
\[ H^0=\C.1,\quad H^1=\C.\theta,\quad H^2=0,\quad
H^3=\C.\star\theta,\quad H^4=\C.{\rm Top}\] and exhibits
Poincar\'e duality.
\end{propos}
\proof Here a closed 0-form means $f$ with $\del^af=0$ for all
$a$, which means $R_a(f)=f$ for all $a$. But $a\in\CC$ genenerate
all of $S_3$ so it means a multiple of $1$. For $H^1$ we consider
a 1-form $A=A^ae_a$ with components $A^a$. Each has six values.
Similarly we take our basis for $\Omega^2$ with
\[
F^{uv}=R_u A^v+A^u-R_w A^u-A^w,\quad F^{vu}=R_v
A^u+A^v-R_uA^w-A^u\]
\[ F^{vw}=R_vA^w+A^v-R_wA^u-A^w,\quad
F^{wv}=R_wA^v+A^w-R_uA^w-A^u\] for the components in our basis.
Hence $\extd$ is an $24\times 18$ matrix \[ \extd_1
=\left(\begin{matrix} \id-R_w& R_u & -\id\\ R_v-\id& \id & -R_u\\
-R_w& \id & R_v-\id \\ -\id& R_w& \id-R_u\end{matrix}\right).\] We
find its kernel, which contains in particular the five
independent exact differentials $\extd \delta_x$ ($x\ne e$, say)
to be six dimensional. Hence $H^1=\C$. It is easy to see that it
is represented by $\theta$ which is closed but not exact. Next,
the image of $\extd$ above must be 12-dimensional. For
$\extd:\Omega^2\to\Omega^3$ we similarly compute
\[ \extd F=(\del^w(F^{uv}-F^{vw})-\del^v F^{wv}+F^{uv}-F^{vu})\h\star e_u
+ (\del^w F^{vu}+\del^u F^{vw}+F^{vu}-F^{uv}+
F^{vw}-F^{wv})\h\star e_v\]
\[ \qquad \qquad +(\del^u (F^{wv}-F^{vu})-\del^v
F^{uv}+F^{wv}-F^{vw})\h\star e_w.\]
We use here
\[ \extd (e_a\wedge e_b)=\h(\star(e_a)-\star(e_b))\]
and the relations in $\Omega^3$. The result can be written as
\[ \extd F=(R_w(F^{uv}-F^{vw})-R_v F^{wv}+F^{vw}+F^{wv}-F^{vu})\h\star e_u
+ (R_w F^{vu}+R_u F^{vw}-F^{uv}-F^{wv})\h\star e_v\]
\[ \qquad \qquad +(R_u (F^{wv}-F^{vu})-R_v
F^{uv}+F^{uv}+F^{vu}-F^{vw})\h\star
e_w\] which is the $18\times 24$ matrix \[ \extd_2=\left(\begin{matrix}
R_w& -\id & \id-R_w &\id-R_v\\-\id& R_w& R_u & -\id\\
\id-R_v& \id-R_u& -\id & R_u\end{matrix}\right)\] which is
basically the transpose of the matrix above for $\extd_1$. Hence
its kernel is 12 dimensional and $H^2=0$. It also means that the
dimension of the space of exact $3$-forms as 12. Next, for $H^4$
we look at $\extd$ on our three-forms. Thus,
\[ \extd\star e_u=2\extd(e_w\wedge e_u\wedge e_v)=2e_u\wedge
e_w\wedge e_u\wedge e_v+2e_w\wedge e_u\wedge e_v\wedge e_u =0\]
hence $\extd f^a\star(e_a)=\del^b e_b\wedge
\star(e_a)=2(\del^af^a){\rm Top}$ is the image of $\extd$ for any
3 functions $f^a$. The $6\times 18$ matrix of $\extd$ on
$(f^u,f^v,f^w)$ is evidently the transpose of the matrix for
$\extd$ on functions, hence its image is 5 dimensional. Note that
this image is precisely the space of functions with zero integral
over $S_3$ (times $\rm Top$). Thus $H^4=\C$ and is represented by
a constant multiple of the top form. Moreover, the kernel of
$\extd:\Omega^3\to\Omega^4$ is therefore 13 dimensional, hence
$H^3=\C$. It is easy to see that it is represented by
$\star\theta$. In particular, we find Poincar\'e duality as
stated. \eproof

\subsection{Spin 1: Maxwell equations}

We now look at the wave operator for spin 1 or
`Maxwell fields' $A$ modulo exact forms.
Here the invariant curvature $F=\extd A$
is a linear version of the true $U(1)$ gauge
theory in the next section. In noncommutative geometry
the latter looks and behaves more like Yang-Mills theory while the
linear theory is more like conventional electromagnetism.

We note that
\[ (\extd A)^*=((R_aA^b+A^a)e_a\wedge e_b)^*=e_b\wedge e_a (R_a
\bar A^b+\bar A^a)=(R_b \bar A^b+R_{ba}A^a)e_b\wedge e_a\qquad\]
\[\qquad\qquad=(A^{*b}+R_bA^{*a})e_b\wedge e_a=\extd (A^*)\]
as it should.
Note that in our basis we have
\[ A^{*a}=R_a(\bar A^a),\quad F^*{}^{ab}=R_{ab}\bar F^{ba}.\]
Then up to total derivatives
\[ L\, {\rm Top}\equiv -{\sqrt{3}\over 4}F^*\wedge\star F
=-{\sqrt{3}\over 4}(\extd A)^*\wedge\star(\extd A)
=-{\sqrt{3}\over 4}A^*\wedge \extd
\star\extd A\] gives the Lagrangian and the required wave
operator \[ \star\extd\star\extd: \Omega^1\to\Omega^1.\] Note
that $\extd(f^a\star(e_a))=2(\del^af^a){\rm Top}$ and $\int
\del^a f^a=0$ means that we can indeed neglect exact 4-forms in
these computations, as we do.

One may also write the Maxwell action more explicitly. Thus \[
{}^\star F^{uv}=-F^{uv}+2F^{vw},\quad {}^\star
F^{vu}=F^{vu}-2F^{wv},\quad {}^\star F^{vw}=F^{vw}-2F^{uv},\quad
{}^\star F^{wv}=-F^{wv}+2F^{vu}\] from which \eqn{LF}{
L=-{1\over4}({\bar{F}^{uv}}(F^{vw}-2F^{uv})+
{\bar{F}^{vu}}(F^{wv}-2F^{vu})+{\bar{F}^{vw}}(F^{uv}-2F^{vw})
+{\bar{F}^{wv}}(F^{vu}-2F^{wv}))} using the relations in
$\Omega^4$ and up to total derivatives. This is
\[
L={1\over 2}(|F^{uv}|^2+|F^{vu}|^2+|F^{vw}|^2+|F^{wv}|^2-{\rm
Re}( \bar F^{uv}F^{vw}+\bar F^{vu}F^{wv}))\] from which the action
is easily seen to be positive semidefinite. Also, it is tempting to
divide $F$
into two halves related through $\star$ much as in the theory of
electromagnetism. One such division is
\[ E=(F^{uv},F^{vu}),\quad B=(F^{vw},F^{wv})\]
since $E,B$ are then rotated componentwise into each other by
$\star$. The action is then the sum of similar parts from $E$ and
from $B$ and a cross term.

\begin{propos} The zero modes of the
wave operator $\star\extd\star\extd$ are precisely the fields of
zero curvature. The equations \[ \extd F=0, \quad \star\extd\star
F=J\] have a solution iff $J$ is `strongly conserved' in the sense
$\extd \star J=0$ and $\int J\wedge\star\theta=0$, and the
solution $F$ is unique. The space of possible sources is
12-dimensional and spanned by four massive $\star\extd\star\extd$
modes for each of the masses $\sqrt{3}$, $\sqrt{6}$ and $3$.
\end{propos}
\proof Putting in the form of $F=\extd A$ into the general
formulae for $\star F$ and $\extd$ on $\Omega^2$ (as given in the
cohomology computation) we obtain $\star\extd\star\extd A$ with
$e_u$ component
\[
R_{uv}A^v+R_{vu}A^w-R_u(A^v+A^w)+R_v(A^u-A^v+A^w)+R_w(A^u+A^v-A^w)
-4A^u+A^v+A^w=0\] and its 2 cyclic rotations. Equivalently, the
matrix for $\star$ on 2-forms in our standard basis is \[
\star_2={1\over\sqrt{3}}\left(\begin{matrix}-1& 0& 2 &0\\0&1&0&-2\\
-2&0&1&0\\0&2&0&-1\end{matrix}\right)\] and as a matrix on the
column vector of the components of $A$,
\[ \star\extd\star\extd=\extd_2\star_2\extd_1={1\over\sqrt{3}}
\left(\begin{matrix}R_v+R_w-4&
R_{uv}-R_u-R_v+R_w+1& R_{vu}-R_u+R_v-R_w+1\\
R_{vu}-R_u-R_v+R_w+1& R_u+R_w-4 & R_{uv}+R_u-R_v-R_w+1\\
R_{uv}-R_w-R_u+R_v+1& R_{vu}-R_w+R_u-R_v+1&
R_v+R_u-4\end{matrix}\right).\] This $18\times 18$ matrix has a
six-dimensional kernel which is the kernel of
$\extd:\Omega^1\to\Omega^2$ as in Proposition~2.3, i.e. it is
precisely the closed forms or forms of zero curvature. It means
that if we solve $\star\extd\star\extd A=J$ for $F$ rather than
for $A$ we have exactly one solution for each $J$ in the image of
the wave operator. The image is therefore 12 dimensional which is
the dimension of the image of $\extd:\Omega^2\to\Omega^3$ in the
cohomology computation, i.e. we require precisely that $\star(J)$
be exact. On the other hand, for any 2-form $F$, $\star\extd F$ as
given in the proof of Proposition~2.3 is such that  $\star\extd
F\wedge\star\theta$ is an exact 4-form. Indeed, its components
are given by adding up the coefficients of $\star e_a$ in
$\star\extd F$, which add up to a total derivative. This
additional property characterises exact 3-forms in the 13
dimensional space of closed 3-forms. Hence in our case $\star J$
exact is therefore characterised by $\extd\star J=0$ and
$J\wedge\star\theta$ an exact 4-form. The latter is the condition
that its integral as a 4-form (which means the usual integral of
the coefficient of $\rm Top$) be zero.

Finally, the other eigenvalues of $\star\extd\star\extd$ are
easily found using the above matrix representation to be $-3,-6$
and $-9$ corresponding to a massive mode as stated. The
application of $\star\extd\star\extd$ to these gives the space of
possible sources. Each eigenspace is 4-dimensional and together
with the zero modes they fully diagonalise $\star\extd\star\extd$.
\eproof

The two conditions for a strongly conserved source can be written
explicitly as
\eqn{scons}{ \sum_a \del^a J^a=0,\quad \int \sum_a J^a=0}
and the second is equivalent to $\sum_a J^a$ a total derivative.
This is stronger than just the usual zero divergence condition
alone precisely due to a nontrivial $H^3$. Other than this
complication (which can arise in the continuum case just as well)
we see that there is a reasonable theory of `electromagnetism' or
`electrostatics'. The explicit form of the equations for $F$ are
the Bianchi equation $\extd F=0$ given explicitly in the proof of
Proposition~2.3 and $\star\extd\star F=J$, which after adding or
subtracting the respective Bianchi identities comes out as
\[ J^u=\del^w F^{uv}-\del^v F^{vu}-F^{vu}+F^{vw}+F^{wv},\quad
J^v=-\del^u F^{uv}-\del^w F^{wv}-F^{vu}-F^{vw}\]
\[ J^w=\del^u F^{wv}-\del^v F^{vw}+F^{uv}+F^{vu}-F^{vw}.\]
And if one want the potential $A$, this is determined only up
to zero modes. These can be gauge fixed by similarly restricting $A$ to
{\em strong Coulomb gauge}
\eqn{coul}{ \sum_a \del^a A^a=0,\quad \int \sum_a A^a=0.}

It remains to construct suitable currents $J$ of a recognizable
form from such a point of view. We obtain them by considering
scalar fields of mass $m$.

\begin{propos} If $\phi$ is an on-shell scalar field of mass $m$ then
\[
J^a=(\del^a\bar\phi)\phi-(R_a\bar\phi)\del^a\phi +{m^2\over
18}\int\bar\phi\phi= 2\del^a(\bar\phi)\phi -
\del^a(\bar\phi\phi)+{m^2\over 18}\int\bar\phi\phi\] is a strongly
conserved current. \end{propos} \proof  Here the `local' term is
obtained by minimal coupling, i.e. from expanding
$((\extd+A)\phi)^*\wedge (\extd+A)\phi$ and has zero divergence.
The $m^2$ term does not change this fact but ensures conservation in our
strong sense.  Thus, from the braided-Leibniz rule we have
\[
\sum_a\del^aJ^a=\sum_a(\del^a\del^a\bar\phi)\phi
 -\sum_a
 \bar\phi\del^a\del^a\phi=-(\square\bar\phi)\phi+\bar\phi(\square\phi)=0\]
when $\phi$ is on shell (an eigenvector of the wave operator). And
\[ \sum J^a=(\square \bar\phi)\phi-\h\square(\bar\phi\phi)+{m^2\over
6}\int\bar\phi\phi\] which has integral zero. The middle term is a
total derivative and does not contribute. \eproof

Hence we have a strongly conserved current for any on-shell
solution $\phi$ of the wave equation. The mass $0,2\sqrt{3}$
solutions from Section~2.1 have zero current. The mass $\sqrt{6}$
modes, however, have a nonzero current. We use the projection given there of
these modes from functions $\phi_0$ and take for
these the `point source' form $\delta_x$. Then the corresponding
`point like' mass $\sqrt{6}$ modes are
\[ \phi=2\delta_x-\delta_{xuv}-\delta_{xvu}.\]
Here
\[ \bar\phi\phi=4\delta_x+\delta_{xuv}+\delta_{xvu},\quad
R_a(\bar\phi)\phi=0\] so that we obtain the current for a `{\em
point-like source}' at $x$, \eqn{pointJ}{
J_x^a=2-R_a(\bar\phi\phi)-\bar\phi\phi=1-3\delta_x-3\delta_{xa}.}
These sources are `radial' in the sense that the component $J^a$
in the $a$ direction of the source located at $x$ has support
along the line $x,xa$ (plus an overall constant value).

These point-like sources at the different $x$ are not independent.
It is easy to see that $J_{xu}+J_{xv}+J_{xw}=0$ so three
point-like sources symmetrically placed about any point cancel out.
Indeed, the above construction gives only 4
independent sources, due to the two relations \eqn{sourctri}{
J_u+J_v+J_w=0,\quad J_e+J_{uv}+J_{vu}=0. }
In fact, these point-like sources span the 4-dimensional $-6$
eigenspace of $\star\extd\star\extd$ which means that the
corresponding potential for a source at $x$ in `strong Coulomb
gauge' is simply
\[A_x=-\frac{1}{6}J_x.\]
Its curvature $F$ may then be
easily computed as \eqn{pointF}{
F^{uv}=\delta_{xu}-\delta_{xw},\quad
F^{vu}=\delta_{xv}-\delta_{xu},\quad
F^{vw}=\delta_{xv}-\delta_{xw},\quad
F^{wv}=\delta_{xw}-\delta_{xu}.}

Next we consider `dipole' configurations. We can clearly polarise
the above formula for $J$ for a scalar field as $J(\phi,\psi)$
where one $\phi$ is replaced by an independent field $\psi$ say.
We still have a strongly conserved source as long as $\phi,\psi$
are on shell with the same mass. Here
\[ J(\phi,\psi)+J(\psi,\phi)=J(\phi+\psi)-J(\phi)-J(\psi)\]
is the source for the combined field minus the source for each
field separately. Letting $\phi,\psi$ be two `point like'
solutions at $x,xb$ respectively (with $b\in \CC$), i.e. a
`dipole' at $x$ in direction $b$, we have $\bar\phi\psi=0$ and
\[
J^a_{x;b}=2R_a(\bar\phi)\psi=(9\delta_{a,b}
-6)(\delta_{xa}+\delta_{xb})+2\sum_{c\in \CC} \delta_{xc}.\]
Here the current is positive when 'lined up' with $b$. This is
our first attempt at a dipole source.
Note that there are only four independent sources
due to the relations:
\eqn{relJdip}{J^a_{x;b}=R^aJ^a_{xb;b},\quad
J^a_{x;u}+J^a_{x;v}+J^a_{x;w}=0,\quad
J^a_{x;b}+J^a_{x(uv);b(uv)}+J^a_{x(uv)^2;b(uv)^2}=0}
and one may find the corresponding potential as
\[A^a_{x;b}=-\frac{1}{9}(2J^a_{x;b}+R^aJ^a_{x;b}).\]
Starting from this source, one can then find nicer formulae if one
introduces a slightly modified source (still satisfying the strong
conservation conditions)
\[J^{\prime a}_{x;b}=J^a_{x;b}+\h\del^a J^a_{x;b}=\h(J^a_{x;b}+J^a_{xb;b})\]
using the first of the relations (\ref{relJdip}). Explicitly
\eqn{dipsource}{J^{\prime a}_{x;b}=1+\h(9\delta_{a,b}-6)(\delta_x
+\delta_{xa}+\delta_{xb}+\delta_{xab}).}
As before there are four
independent configurations here and they span the eigenspace of
$\star\extd\star\extd$, now of eiganvalue -3. The corresponding dipole
potential is therefore
\[A^a_{x;b}=-\frac{1}{3}J^{\prime a}_{x;b}.\]
Its curvature can easily be computed and one finds for a dipole
centered at $x$ and directed along $b=u$ (say), \eqn{Fdipa}{
F^{uv}_{x;u}=9(\delta_{xu}-\delta_{xvu}+\delta_x-\delta_{xw}),\quad
F^{vu}_{x;u}=9 (\delta_{xuv}-\delta_{xu}+\delta_{xv}-\delta_x)}
\eqn{Fdipb}{
F^{vw}_{x;u}=9(-\delta_{xvu}-\delta_{xv}-\delta_{xuv}-\delta_{xw}
-\delta_{xuv}-\delta_{xv}),\quad
F^{wv}_{x;u}=9(\delta_{xu}-\delta_{xvu}+\delta_{xw}-\delta_{x}
-\delta_{xw}-\delta_{xvu}).
} This gives an electrostatics picture of some of the massive spin
one modes. Note that the mass here, as for the lower spins,
reflects the background constant curvature of $S_3$ in the sense
of \cite{Ma:fin}.

\section{U(1) noncommutative Yang-Mills theory} \label{U(1) noncomm}

Here we do $U(1)$ `gauge theory' in the more usual sense. In usual
commutative geometry this essentially coincides with cohomology
theory but in the noncommutative case the curvature \eqn{FAA}{
F=\extd A+A\wedge A} remains nonlinear. It is covariant as
$F\mapsto UFU^{-1}$ under \eqn{AU}{ A\mapsto UAU^{-1}+U\extd
U^{-1},\quad A^a \mapsto {U\over R_a(U)}A^a+U\del^a U^{-1}} for
any unitary $U$ (i.e. any function of modulus 1). Here we limit
attention to `real' $A$ in the sense $A^*=A$. This translates in
terms of components as
\[ \bar A^a=R_a A^a,\quad \bar F^{ab}=R_{ab}(F^{ba})\]
and implies that $F^*=F$ is `real'.

Our first step is to change variables to $A=\Phi-\theta$, i.e.
$A^a=\Phi^a-1$ and certain operators $\rho_a\equiv \Phi^a R_a$
\eqn{FYMzeroa}{ F^{uv}=\rho_u\Phi^v-\rho_w\Phi^u,\quad
F^{vu}=\rho_v\Phi^u-\rho_u\Phi^w} \eqn{FYMzerob}{
F^{vw}=\rho_v\Phi^w-\rho_w\Phi^u,\quad
F^{wv}=\rho_w\Phi^v-\rho_u\Phi^w.} Here $\Phi^a\mapsto {U\over R_a
U}\Phi^a$ transforms covariantly and \eqn{realPhi}{
\bar\Phi^a=R_a\Phi^a} is our reality constraint. The reality
constraint means that $\Phi^a$ are determined freely by their
values on $u,v,w$. It also means that
\eqn{lambda}{\lambda_a^2\equiv |\Phi^a|^2} are real-valued
gauge-invariant function associated to any gauge field.

\subsection{Zero curvature moduli space}

In classical geometry the zero curvature gauge fields detect the
`homotopy' or fundamental group of a manifold. Hence in
noncommutative geometry the presence of a moduli of flat
connections is indicative of this.  We find it to be nontrivial.

\begin{theorem} The moduli space of zero curvature gauge fields
modulo gauge transformation is the union of a 1-parameter
positive half-line
\[A=(\mu-1)\theta,\quad \mu\ge 0\]
and six positive cones of $\R^3$ of the form
\[ A=\Phi-\theta,\quad \Phi^a(b)=\mu^a{}_b,\quad a,b\in \CC\]
where $\mu^a{}_b\ge 0$ are a matrix of the form
\[{\rm (i):}\  \left(\begin{matrix}*&*&*\\
0&0&0\\0&0&0\end{matrix}\right),\quad\left(\begin{matrix}0&0&0\\
*&*&*\\0&0&0\end{matrix}\right),\quad\left(\begin{matrix}0&0&0\\
0&0&0\\ *&*&*\end{matrix}\right),\ {\rm (ii):}\ \left(\begin{matrix}*&0&0\\
0&0&*\\ 0&*&0\end{matrix}\right),\quad\left(\begin{matrix}0&0&*\\
0&*&0\\ *&0&0\end{matrix}\right),\quad\left(\begin{matrix}0&*&0\\
*&0&0\\0&0&*\end{matrix}\right).\]
\end{theorem}
\proof Given any zero curvature solution we clearly have
 \[ \rho_u \Phi^v=\rho_w\Phi^u=\rho_v\Phi^w, \quad
 \rho_u\Phi^w=\rho_v\Phi^u=\rho_w\Phi^v.\]
In fact these two equations are equivalent under the reality assumption but
it is useful to work with both forms. Then
\[ \rho_u\rho_u\Phi^v=\rho_u(\Phi^u R_u\Phi^v)=\Phi^v\lambda_u^2
=\rho_u\rho_w\Phi^u=\rho_u(\Phi^w)R_{vu}(\Phi^u)=\Phi^vR_v(\lambda_u^2)\]
\[
\quad\quad=\rho_u\rho_v\Phi^w=\rho_u(\Phi^v)R_{uv}\Phi^w
=\rho_v(\Phi^w)R_{vw}(\Phi^w)
=\Phi^vR_v(\lambda_w^2).\] Hence
\[ \lambda_v^2(\lambda_w^2-\lambda_u^2)=0,\quad
\lambda_v^2\del^v\lambda_u^2=0,\quad
\lambda_v^2\del^v(\lambda_w^2)=0\] and the cyclic rotations of
these. We also have $\del^v\lambda^2_v=0$, etc. Choose any point
(for a nonzero configuration) where $\Phi^v(x)\ne 0$, say. Then
$\lambda_u(x)= \lambda_w(x)$ so either both are zero or not.
Assume the latter (so all components at $x$ are nonzero). Then
$\lambda_u(xv)=\lambda_w(xv)\ne 0$ since $\lambda_v(x)\ne 0$, and
$\lambda_v(xv)=\lambda_w(xv)$ since $\lambda_u(xv)\ne 0$, hence
all components at $xv$ are also nonzero. Iterating, we conclude
in this case that $\lambda^2_u=\lambda^2_v=\lambda^2_w=\mu^2$
say, where $\mu$ is a positive constant. The other possibility is
that at every $x\in S_3$ at most one component is nonzero, which
degenerate case will be handled later.

In the nowhere zero case, we consider the gauge transform
\[U(e)=1,\quad U(u)={\Phi^u(e)\over\lambda},\quad U(v)
={\Phi^v(e)\over\lambda},
\quad U(w)={\Phi^w(e)\over\lambda},\]
\[ U(uv)={\Phi^u(e)\Phi^v(u)\over\lambda^2},\quad
U(vu)={\Phi^w(e)\Phi^v(w)\over\lambda^2}\] which is manifestly
unitary. Using the zero curvature conditions one may check that
indeed it gauge transforms $\Phi$ to
$\Phi^u=\Phi^v=\Phi^w=\lambda$.

We turn now to the degenerate case where at each point at most one
component of $\Phi$ is nonzero. Note first that we need only be
concerned with the matrix $\{\Phi^a(b)\}$ where $a,b$ run over
$u,v,w$, since the reality condition determines the values then
at $e,uv,vu$. Moreover, the reality condition then becomes empty.
For example, $\Phi^u(uv)=\Phi^u(wu)=\bar\phi^u(w)$ and
$\Phi^v(uv)=\bar\Phi^v(u)$, etc. Next, under a gauge transform
this matrix goes to
\[ \Phi^a{}_b\to\Phi^a(b){U(b)\over U(ba)}\]
and because under our degeneracy assumption of at most one entry
in each column is nonzero, we can chose this in such a way that
all nonzero entries can be gauge transformed onto
 the real positive axis. Indeed, we chose
\[ U(e)=U(uv)=U(vu)=1,\quad U(b)={|\Phi^a(b)|\over \Phi^a(b)}\]
where there is at most one nonzero $\Phi^a(b)$ at each $b=u,v,w$
(and we set $U(b)=1$ if there is none). Thus every zero curvature
solution of our degenerate type is gauge equivalent to one where
the matrix is given by real non-negative numbers $\mu^a{}_b$ of at
most 3 entries. These are equal to the gauge invariant norms
$\lambda^2_a$ and cannot be transformed further while remaining
on the positive real axis, so there is one solution for each
allowed matrix.

Precisely which matrices are allowed is determined by the
zero-curvature equation. Writing this out in terms of the
$\Phi^a(b)$ we have
\[ \Phi^u(u) \Phi^v(v)=\Phi^v(u) \Phi^w(v)=\Phi^w(u) \Phi^u(v),\quad
\Phi^u(v) \Phi^v(w)=\Phi^v(v)\Phi^w(w)=\phi^w(v)\Phi^u(w)\]
\[ \Phi^u(w)\Phi^v(u)=\Phi^v(w)\Phi^w(u)=\Phi^w(w)\Phi^u(u)\]
for the zero curvature at $u,v,w$. At the other points it yields
\[ \Phi^v(u)\Phi^u(u)=\Phi^w(v)\Phi^v(v)=\Phi^u(w)\Phi^w(w),\quad
\Phi^v(w)\Phi^u(w)=\Phi^w(u)\Phi^v(u)=\phi^u(v)\Phi^w(v)\]
\[ \Phi^v(v)\Phi^u(v)=\Phi^w(w)\Phi^v(w)=\Phi^u(u)\Phi^w(u)\]
which is empty in our case where every column has at most one
nonzero entry (it is the origin of this restriction).

Finally, we enumerate the allowed patterns. (i) Clearly if two
rows (i.e. two of the $\Phi^u,\Phi^v,\Phi^w$ are entirely zero)
then the third is free for a zero curvature solution. This is the
first set of matrices shown. (ii) If exactly one row is entirely
zero, say $\Phi^w$, then the other two obey
\[ \Phi^u(u)\Phi^v(v)=0,\quad \Phi^u(v)\Phi^v(w)=0,
\quad\Phi^u(w)\Phi^v(u)=0\]
from the first zero of zero curvature equations. This says that
the $\Phi^u$ row has no nonzero entries with the rotated $\Phi^v$
row. If one row has more than one nonzero entry then this forces
the other row to be entirely zero as well and we are back in case
(i). Otherwise neither row can can have more than one nonzero
entry which means that we are either in case (i) again or in a
degenerate case of the next case. (iii) The remaining case is when
each row $\Phi^a$ has at most one nonzero entry
$\Phi^a(\sigma(a))$, say, for some permutation $\sigma$ of $u,v,w$
(anything else would imply one of the rows was entirely zero,
covered above). In this case we have potentially 6 possibilities
depending on $\sigma\in S_3$. Now, for this type of solution the
zero curvature equation reads
\[ \Phi^a(\sigma(a))\Phi^b(\sigma(b))=0,\quad {\rm if}\
\sigma(a)\sigma(b)=ab.\]
For $\sigma=\id$ this means $\Phi^a(a)\Phi^b(b)=0$ for all $a,b$,
which means that two out of three of our rows must be zero, which
puts us back in case (i) above. Similarly if $\sigma$ is a
rotation $u\to v\to w\to u$ or its inverse then we have three
equations forcing two out of three to be zero and hence in case
(i). The three remaining possibilities are where $\sigma$ fixes
one of $u,v,w$ and flips the other two. In this case the relations
are empty i.e. we can freely chose the potentially nonzero matrix
entries $\Phi^a(\sigma(a))$. This is the second family of
positive cones in $\R^3$ stated. Note that the matrices of
$u,v,w$ themselves in their natural representation on 3 elements
are in this second family. \eproof

Similarly, in terms of the components of $F$ and $\star{F}$ as in
the previous section, we have the self-duality equation as
\[ F^{vw}=\lambda F^{uv},\quad F^{wv}=\lambda^{-1}F^{vu},\quad
\lambda=\h(1+\imath\sqrt{3})\] after collecting terms. Note that
$|\lambda|=1$ and $\lambda^3=-1$. Under our reality condition
only one of these equations is needed, the other being
equivalent. We see that a selfdual 2-form subject to our reality
condition is therefore determined entirely by an unconstrained
complex function $F^{uv}$.

One could therefore ask for the moduli of self-dual gauge fields
or `instantons', i.e. when such 2-forms can be the curvature of a
gauge field. Note that there can be no self-dual Maxwell
connections other than $F=0$ due to the unique solution of the
Maxwell equations for $F$ with no source (as seen in the preceding
section). Therefore one should not necessarily expect instantons
here either. Indeed, putting in the form of $F$ for the $U(1)$
Yang-Mills theory, we obtain the self-duality equations as
\[ \rho_u\Phi^v=\lambda^{-1}\rho_v\Phi^w+\lambda\rho_w\Phi^u \]
and our `reality' constraint on the $\Phi$. This appears to have
no solutions.

\subsection{Yang-Mills action and other extrema}

Finally, we take a look at the Yang-Mills action in general.  In
terms of $F$ the Lagrangian is exactly the same as that stated in
Section~2.4 for the Maxwell field, and is therefore positive
semidefinite. In our Yang-Mills case we put in the form of $F$ in
terms of $\Phi$.

\begin{theorem} The (rescaled) Yang-Mills action in terms of the
gauge field fluctuation $\Phi$ and up to total derivatives is
\[ L=-{\sqrt{3}\over 4}F^*\wedge\star F
=\lambda^2_u R_u \lambda^2_v-\Phi^u
(R_u\Phi^v)(R_{uv}\Phi^u)(R_w\Phi^w)+{\rm
cyclic}\] and is positive semidefinite.
\end{theorem}
\proof We put the form of $F$ into the second expression for the
Lagrangian in Section~2.4. First we explicitly put in the reality
condition on the $F$ which implies that \[
L=|F^{uv}|^2+|F^{vw}|^2-{\rm Re}(\bar F^{uv}F^{vw})\] up to a
total derivative. Then
\[ |F^{uv}|^2=R_{uv}(\Phi^v R_v\Phi^u-\Phi^u R_u\Phi^w)(\Phi^u
R_u\Phi^v-\Phi^w R_w\Phi^u)\] \[\qquad=\lambda^2_u
R_u\lambda^2_v+\lambda^2_w R_w\lambda^2_u-2\Phi^w
(R_w\Phi^u)(R_u\Phi^u)R_{uv}\Phi^v\] up to a total derivative.
Similarly \[ |F^{vw}|^2=\lambda^2_v R_v\lambda^2_w +\lambda_w^2
R_w\lambda^2_u-2\Phi^w (R_v\Phi^v)(R_w\Phi^u)R_{uv}\Phi^w.\]
Finally, we compute
\[
{\bar{F}^{uv}}F^{vw}=(R_u\Phi^u)\Phi^v(R_v\Phi^v)R_{uv}\Phi^v+\lambda^2_w
R_w\lambda^2_u-(R_w\Phi^w)\Phi^v(R_v\Phi^w)R_{uv}
\Phi^u-(R_u\Phi^u)\Phi^w(R_w\Phi^u)R_{uv}\Phi^v.\]
Adding the minus the real part of this to the other terms and
discarding total derivatives gives the result for $L$. \eproof

From the physical point of view this result is very significant.
It states that when we write the values of $\Phi^a(x)$ in polar
coordinates their gauge-invariant fields $\lambda^2_a(x)$
contribute like some kind of massive particle with Lagrangian
\eqn{L0}{ L_0=
\lambda^2_u\del^u\lambda^2_v+\lambda^2_v\del^v\lambda^2_w
+\lambda^2_w\del^w\lambda^2_u +\lambda^2_u\lambda^2_v
+\lambda^2_v\lambda^2_w+\lambda^2_w\lambda^2_u} and a part given
by the sum of the Wilson loops $W_u$, $W_v$, $W_w$ at $x$. Here
\eqn{Wu}{ W_u=\Phi^u (R_u\Phi^v)(R_{uv}\Phi^u)(R_w\Phi^w),\quad
W_u(x)=\Phi^u(x)\Phi^v(xu)\Phi^u(xuv)\Phi^w(xw)} is the product
around a path defined by right translating by $u$, then by $v$
then by $u$ ($=u^{-1}$) and then by $w$. Here $uvuw=e$ is a
relation in $S_3$ in terms of our elements of $\CC$ and such
relations form our elementary plaquettes. One can also introduce
homology and homotopy of allowed paths in the group as defined by
filling in via elementary plaquettes, i.e. one should think of
them as `pieces of area' defined by the differential calculus.

We will say more about the $U(1)$ lattice gauge theory defined by
the angular part of $\Phi=\lambda e^{\imath\theta}$ in the next
section. At present we concentrate on the real-positive radial
variables $\lambda$ with `free particle' Lagrangian $L_0(\lambda)$
(which is quadratic in terms of the functions $\lambda_a^2$). Note
that $\lambda_a(x)=\lambda_a(xa)$, i.e. these variables are really
associated to the steps (edges) along allowed directions in the lattice.
They are a hybrid of some kind of `length' or `metric'  assignment
 to the abstract lattice on which the more conventional
$U(1)$ gauge theory takes place, and the real part of the field strength
of $A$ (they are the modulus of the infinitesimal transport `$1+A^a(x) \extd x_a$'
and hence involve both features rolled into one.) The noncommutative
Yang-Mills theory factorises into some kind of `metric' theory
for the $\lambda$ and a conventional lattice $U(1)$ for the angular
variables. Apart from $L_0$ there is also an interaction term
coming from the polar decomposition
\[
W_u(x)=\lambda_u(x)\lambda_v(xu)\lambda_u(xuv)\lambda_w(xw)w_u(x)\]
where $w_u$ (etc.) are the conventional $U(1)$-valued Wilson
loops. One may heuristically think of expressions such as $\lambda_u^2\lambda_v^2$
in $L_0$ as `area' of an elementary plaquette and the products of the
$\lambda$ in the $W_u$ as `multiplicative perimiter'. It is interesting that
both expressions are quartic, which is consistent with the idea that
holonomies in finite lattice theory go as area law (this would becomes Wilson's
criterion for confinement if it survived to the continuum limit, but we
are not able to consider this in our finite model). Note also that
a flat connection $A$ corresponds to both a flat $U(1)$
connection in the sense of trivial holonomy around the elementary
plaquettes as above {\em and} a flat assignment of the $\lambda$
variables when multiplied. The physical meaning of this is not clear
(it comes from the field strength nature of the $\lambda$ and perhaps
suggests to think of them as transition
probabilities when suitably normalised). At any rate, one has a real $\R_+$-valued
gauge theory for the $\lambda$ in the finite geometry as well as a $U(1)$ lattice
theory. These are quite general features that apply
for other groups also.

In particular, we can look at the pure `metric'  sector of the theory
where all the $U(1)$-Wilson loops $w_a$ are constrained to be $1$.
For example we can take all the $\Phi$ real. In any case the only variables
entering are then the $\lambda$ and the total action in terms of the
nine variables $\{\lambda_a(b)\}$ assigned to the link $a,ab$ is
\[ S=\int L=\lambda^2_u(u)\lambda^2_v(u)+\lambda^2_u(u)\lambda^2_v(v)+
\lambda^2_u(v)\lambda^2_v(w)+\lambda^2_u(w)\lambda^2_v(u)+\lambda^2_u(w)
\lambda^2_v(w)+\lambda^2_u(v)
\lambda^2_v(v)\qquad\]
\eqn{Sl}{
-2\lambda_u(u)\lambda_v(v)\lambda_u(v)\lambda_w(u)-2\lambda_u(v)
\lambda_v(w)\lambda_u(w)\lambda_w(v)-2\lambda_u(w)\lambda_v(u)
\lambda_u(u)\lambda_w(w)}
plus the cyclic rotations $u\to v\to w\to v$ of all terms. The first
set of terms
(which are
$\int L_0$) can be written as a symmetric quadratic form $D$ on the vector
$(\lambda^2_u(u),\lambda^2_u(v),\lambda^2_u(w),\lambda^2_v(u),\cdots,
\lambda^2_w(w))$. Here
\eqn{metD}{D=\h \left(\begin{matrix}0& 0& 0& 1& 1& 0& 1& 0& 1
\\ 0& 0& 0& 0& 1& 1& 1& 1& 0\\ 0& 0& 0& 1& 0& 1&
    0& 1& 1\\ 1& 0& 1& 0& 0& 0& 1& 1& 0\\ 1& 1& 0& 0& 0& 0& 0& 1& 1\\ 0& 1&
     1& 0& 0& 0& 1& 0& 1\\ 1& 1& 0& 1& 0& 1& 0& 0& 0\\ 0& 1& 1& 1& 1& 0& 0&
    0& 0\\ 1& 0& 1& 0& 1& 1& 0& 0& 0\end{matrix}\right)}
is diagnonalisable over $\R$ and has 4 eigenvectors of
eigenvalue $-1$ and 4 of eigenvalue $1/2$. There is a final mode
of eigenvalue $2$ which is the vector with all entries $\lambda=1$,
which corresponds to
$A=0$. It corresponds to an equal length for all allowed directions.
Because all the eigenmodes are real, we can linearise the theory about this
configuration and our positivity constraints are not affected. On the
other hand, the $\lambda=1$ solution is an absolute minimum of
the total action $S$ (using Theorem~3.2). Hence all these fluctuations
increase the energy of the configuration. In particular, do
not appear to have `metric waves' in the theory for this model.
Theorem 3.1 tells us that there are other 3-manifolds of flat connections
in families (i), (ii) which are singular in the sense that some
$\lambda_a(b)$ vanish. Their fluctuations (by the theorem) have
three modes which keep the action zero but for which the connection remains
singular, while other fluctuations increase the action. It  appears
 from this
discussion (without actually trying to do the integrals)
that the `quantum statistical mechanics' of  this
theory (i.e integrals over the nine $\lambda$ variables with weighting
$e^{-S}$) has $ <\lambda^a(b)>>0$. Note also that this `metric' theory
of these $\lambda$ should not, however, be confused with the actual
noncommutative
Riemannian geometry as in \cite{Ma:fin} which is based on spin connections
rather than $U(1)$ connections $A$, but
it gives some flavour of the full theory.

\section{Quantum electromagnetism}

In this section we conclude with some basic aspects of the
formulation of the quantum theory using a path integral approach.
We will show that the quantum theory is fully computable. Indeed,
functional integration in our setting becomes finite-dimensional
iterated integrals and one can in fact do these integrals. For the
present we also omit physical constants and factors of $\imath$ in
the action since these are matter of taste. Since there is no
preferred time direction one might think that the Euclidean theory
is more appropriate.

We begin with the simplest case, a free scalar field.
\[ Z_A=\int D\phi\ e^{\int (\extd\phi)^*\wedge\star\extd\phi
+V(\phi)+\int A^*\wedge\star J(\phi)}\] for some potential
$V(\phi)$ and possible coupling to an external field $A$. On
shell, the current $J$ is conserved but one should not exactly
think that $A$ is a Maxwell field. For a more geometrical theory
of a particle moving in a background potential one should use the
gauge theory and minimal coupling method (see below). The main
feature of the above is that it is fully computable by elementary
means, depending on the potential and external field. Of course
there is nothing stopping one doing some of these functional
integrals (and those below) using Feynman diagram methods and a
perturbative approach, which may be useful (depending on the
potential $V$).

Equally elementary, we can quantise the Maxwell field with a
classical external source $J$. Thus,
\[ Z[J]=\int DA\  e^{\int (\extd A)^*\wedge \star (\extd A)
+ A^*\wedge\star J}\]
where we have a infinite gauge degeneracy. This can be handled in
several ways. For example we regularise integrals to a finite
volume of field strength of modulus $< \Lambda$, and take
$\Lambda$ to infinity. Gauge symmetry means a factor $\Lambda^6$
but in the ratios involved in vacuum expectation values this
cancels, i.e. we can regulate and remove regulator in all ratios
with ease. More geometrically, we have already seen that the
strengthened Coulomb gauge (\ref{coul}) in Section ~2.4 is a complete gauge
fixing. Hence we can impose these by integrating over a
functional Lagrange multiplier  field (Faddeev-Popov ghosts) for
the $\del^a A^a=0$ condition, and an additional constant Lagrange
multiplier for the global $\int\sum_a A^a=0$ condition.

On the other hand, neither of these conventional formalities are
needed in our finite case. This is because we know that the operator
$\star\extd\star\extd$ in Section~2.4, while not symmetric,
can be diagonalised via Gram-Schmidt to orthonormal eigenvectors $e_i$ say,
$i=1,\cdots, 12$ for the 12-dimensional space of nonzero
eigenvalue. Being eigenvectors these are also in the image of the
operator and can therefore be viewed either as strongly conserved
sources $J$ or gauge potentials $A$ in the strong Coulomb gauge.
We have seen in our case that there are 4 eigenvectors each of
eigenvalue $-3,-6,-9$. Clearly, if we write
$A=\alpha^i e_i$ and $J=J^i e_i$ and the eigenvalues are
$\lambda_i$  then
\[ Z[J]=\int \extd^{12} \alpha^i\
e^{2\lambda_i|\alpha_i|^2+2\bar\alpha_iJ^i}.\] We need here that
$A^*\wedge\star J=A^*{}^ae_a\wedge J^b\star e_b=2R_a(\bar A^a
J^a){\rm Top}$ so that its integral is the usual $l^2$ inner
product on $S_3$.

For a less trivial theory one can also couple the two theories
above, thus
\[ L=(\extd\phi)^*\wedge(\extd\phi)+(\extd A)^*\wedge \star
(\extd A) + A^*\wedge J(\phi).\] This is not gauge invariant
(except when $\phi$ is on shell) but it can still be functionally
integrated over.

Finally, and more interesting than the essentially linear or
Maxwell theory is the fully nonlinear Yang-Mills theory even in
the $U(1)$ case. Here we have been rather more careful to impose
the unitarity condition (because it has more of an impact) in our
treatment in Section~3. In particular, we really do not need to
gauge fix since the $U(1)^6$ symmetry gives a finite volume
$(2\pi)^6$.

Similarly in this case there is a covariant derivative under a
gauge symmetry $\phi\mapsto U\phi$ for charged scalar fields (for
example),
\[ D_A\phi=(\extd+A)\phi,\quad D_A\phi\mapsto
\extd(U\phi)+(UAU^{-1}+U\extd U^{-1})U\phi=UD_A\phi.\] Then
\[ L=F^*\wedge\star F+(D_A\phi)^*\wedge\star D_A\phi+V(\bar\phi\phi)\]
is the Lagrangian for the coupled system with some potential $V$.
We have used part of this for the source $J(\phi)$ and this is its
proper context.
Of particular interest is the pure Yang-Mills theory. In lattice
gauge theory, even for $U(1)$ one expects confinement as an
artefact of the lattice regularisation. In our noncommutative
geometrical version of lattice theory this appears as the
$A\wedge A$ term which does not vanish precisely because the
differential calculus is noncommutative. This it enters in the
same `form' as in nonAbelian gauge theory but for a different
reason, but one may logically expect similar behaviour. Here we
only want to note that our elementary `Wilson loops' are in fact
gauge invariant and our result in Theorem~3.2 for the form of
their action makes it particularly easy to compute them as
follows. We define
\[ Z[\mu_u,\mu_v,\mu_w]=\int\extd A\ e^{\int L_0-\mu_u W_u-\mu_v
W_v-\mu_w W_w}\] where $L_0$ is the $\lambda_a$ part of
the Lagrangian given in (\ref{L0}). We can then compute the
expectation values of elementary Wilson loops as
\[
\<W_u(x)\>=-Z^{-1}{\delta\over\delta\mu_a(x)}|_{\mu_u=\mu_v=\mu_w=1}
(Z).\] This is a matter of 9 complex or 18 real integrals for the
fields $\Phi^a(b)$ which determine the gauge configuration
$A=\Phi-\theta$ (as explained in the proof of Theorem~3.1). We
compute the detailed form of the theory now (actual numerical
computations will be attempted elsewhere).

Thus, given the nine $\Phi^a(b)$ for $a,b\in \CC$, the other
$\Phi^a(x)$ are determined by the reality conditions, so we have
only to  integrates over all the possibles complex values for
these nine. Next we adopt polar coordinates as in Theorem~3.2,
\[\Phi^a(b)=\lambda_a(b) e^{\imath \theta^a(b)},\quad
\lambda_a(b)\in[0,\infty),\quad \theta^a(b)\in [0,2\pi).\]
Including the Jacobian determinant, the partition function becomes
\[Z=2^{-9}\int_0^{\infty} \extd^9\lambda^2\  e^{\int L_0(\lambda)}
\int_{0}^{2 \pi} \extd^9\theta\
\ e^{ - \int W_u +W_v+W_w}\] Here
$\extd^9\lambda^2=\extd \lambda^2_u(u)\cdots\extd \lambda^2_w(w)$ as
in Section~3.2 and $\extd^9\theta=\extd\theta^u(u)\cdots\extd\theta^w(w)$.
We omit the $\mu$ for simplicity. Next we write the Lagrangian in
this integral explicitly in terms of these variables. Thus
\[\h\int W_u +W_v+W_w=\lambda^u(u)\lambda^v(v)\lambda^u(v)
\lambda^w(u) \cos(\theta^u(u)-\theta^v(v)+\theta^u(v)-\theta^w(u))\]
\[\qquad\qquad\qquad+\lambda^u(v)\lambda^v(w)\lambda^u(w)\lambda^w(v)
\cos(\theta^u(v)-\theta^v(w)+\theta^u(w)-\theta^w(v))\]
\[\qquad\qquad\qquad+\lambda^u(w)\lambda^v(u)\lambda^u(u)\lambda^w(w)
\cos(\theta^u(w)-\theta^v(u)+\theta^u(u)-\theta^w(w))\]
plus the cyclic rotations $u\to v\to w\to u$.

We concentrate on the $\theta$-integrals, i.e. we write
\[ Z=\int_0^\infty \extd^9 \lambda^2\ e^{\int L_0(\lambda)}\ Z_\lambda\]
where $Z_\lambda$ is the partition function for the $U(1)$ lattice gauge theory
defined by the $\theta$ variables with the $\lambda$ variables held fixed.
Next, gauge symmetry means that the Lagrangian here does not in fact depend on
all nine of the $\theta$ parameters. In fact it depends on only four, which can be
made manifest
by the transformation matrix:
\[\left(
\begin{array}{ccccccccc}
0& 1& 1& 0& 0& -1& 0& -1& 0\\
-1& 0& 0& 1& 1& 0& 0& -1& 0\\
0& 0& -1& 1& 0& 1& -1& 0& 0\\
0& 0& -1& 0& -1& 0& 0& 1& 1\\
0& 0& 0& 0& 1& 0& 0& 0& 0\\
0& 0& 0& 0& 0& 1& 0& 0& 0\\
0& 0& 0& 0& 0& 0& 1& 0& 0\\
0& 0& 0& 0& 0& 0& 0& 1& 0\\
0& 0& 0& 0& 0& 0& 0& 0& 1\\
\end{array}
\right)
\]
Explicitly, we replace the $\{\theta^a(b)\}$ by
\[\begin{array}{l}
\theta_1=\theta^u(v)-\theta^v(w)+\theta^u(w)-\theta^w(v)\\
\theta_2=\theta^v(u)-\theta^w(v)+\theta^v(v)-\theta^u(u)\\
\theta_3=\theta^v(w)-\theta^w(u)+\theta^v(u)-\theta^u(w)\\
\theta_4=\theta^w(v)-\theta^u(w)+\theta^w(w)-\theta^v(v)\\
\end{array}
\]
and the remaining five
\[ \theta_5=\theta^v(v),\quad
\theta_6=\theta^v(w),\quad
\theta_7=\theta^w(u),\quad
\theta_8=\theta^w(v),\quad
\theta_9=\theta^w(w)\]
are unchanged. The determinant for this change of variables is $1$. We also
write
\[
\lambda_1=\lambda_u(u)\lambda_v(v)\lambda_u(v)\lambda_w(u),\quad
\lambda_2=\lambda_u(v)\lambda_v(w)\lambda_u(w)\lambda_w(v),\quad
\lambda_3=\lambda_u(w)\lambda_v(u)\lambda_u(u)\lambda_w(w)\]
for the $\lambda$-holonomy expressions as in (\ref{Sl}). Similarly
\[\lambda_4=\lambda_v(u)\lambda_w(v)\lambda_v(v)\lambda_u(u),\quad
\lambda_5=\lambda_v(v)\lambda_w(w)\lambda_v(w)\lambda_u(v),\quad
\lambda_6=\lambda_v(w)\lambda_w(u)\lambda_v(u)\lambda_u(w)\]
\[\lambda_7=\lambda_w(u)\lambda_u(v)\lambda_w(v)\lambda_v(u),\quad
\lambda_8=\lambda_w(v)\lambda_u(w)\lambda_w(w)\lambda_v(v),\quad
\lambda_9=\lambda_w(w)\lambda_u(u)\lambda_w(u)\lambda_v(w)\]
for their cyclic rotations. Then we arrive at our final result \eqn{Zl}{
Z_\lambda=\int_0^{2\pi}\extd\theta_5\cdots\extd\theta_9
\int_D\extd\theta_1\cdots\extd\theta_4\
e^{-S_\lambda(\theta_1,\theta_2,\theta_3,\theta_4)},} where
\[\h S_\lambda=\lambda_1 \cos(\theta_1-\theta_2+\theta_3)+\lambda_2
\cos(\theta_1)+\lambda_3 \cos (-\theta_2-\theta_4)+
\lambda_4 \cos (\theta_2)+\lambda_5 \cos(-\theta_1-\theta_4)\]
\[\qquad+\lambda_6 \cos(\theta_3)+ \lambda_7
\cos(-\theta_1-\theta_3)+\lambda_8\cos(\theta_4)+ \lambda_9
\cos(\theta_2-\theta_3+\theta_4)\] and where the domain $D$ is an
affine transformation in $\R^4$ of the hypercube $[0,2\pi)^4$.
That is, it has the form
\[ \left(\begin{array}{c}\theta_1\\ \theta_2\\ \theta_3\\ \theta_4\end{array}\right)
=M\left([0,2\pi)^4\right)+\left(\begin{array}{c}c_1\\ c_2\\ c_3\\
c_4\end{array}\right) ,\] where the linear transformation of the
hypercube is given by \eqn{M}{M=\left(
\begin{array}{cccc}
0& 1& 1& 0 \\
-1& 0& 0& 1 \\
0& 0& -1& 1 \\
0& 0& -1& 0 \\
\end{array}
\right)} and the offsets (which are the only parts that depend on
$\theta_5,\cdots,\theta_9)$ are \eqn{c}{
c_1=-\theta_6-\theta_8,\quad c_2=\theta_5-\theta_8,\quad
c_3=\theta_6-\theta_7,\quad c_4=-\theta_5+\theta_8+\theta_9.}
Clearly, one may compute the domain of integration
$M([0,2\pi)^4)$  for the variables $\theta'_i=\theta_i-c_i$ and
thereby do the four $\theta'$ integrations followed by more
trivial $\theta_5,\cdots,\theta_9$ integrals. Without doing the
actual integrals, it is clear at this point that one obtains here
some form of Bessel function (if we put an $\imath$ in the action)
as $\int_0^{2\pi}\extd\theta
e^{\imath\lambda\cos\theta}=2 \pi J_0(\lambda)$. Similarly higher Bessel
functions for expectation values of the $U(1)$ Wilson loops
$w_u(x)$, etc. This is a similar situation as conventional lattice
gauge theory. In addition we have the `metric' $\lambda$ integrals
in our theory as discussed in Section~3.2.

\subsection*{Acknowledgements} We would like to thank
Xavier Gomez for some useful comments.

%\bibliographystyle{unsrt}
%\bibliography{biblio}

\end{document}